# Consequences of breaking time reversal symmetry in LaSb: a resistivity plateau and extreme magnetoresistance

F. F. Tafti [1*], Q. D. Gibson [1], S. K. Kushwaha [1], N. Haldolaarachchige [1], and R. J. Cava [1*]

1. Department of Chemistry, Princeton University, Princeton, New Jersey, 08544, USA

*Corresponding author: F. F. Tafti (ftafti@princeton.edu)

Time reversal symmetry (TRS) protects the metallic surface modes of topological insulators (TIs).[1] The transport signature of robust metallic surface modes of TIs is a plateau that arrests the exponential divergence of the insulating bulk with decreasing temperature. This universal behavior is observed in all TI candidates ranging from $Bi_2Te_2Se$ (BTS) to $SmB_6$.[2–5] Recently, several topological semimetals (TSMs) have been found that exhibit extreme magnetoresistance (XMR) and TI universal resistivity behavior revealed only when *breaking* TRS, a regime where TIs theoretically cease to exist. Among these new materials, TaAs and NbP are nominated for Weyl semimetal due to their lack of inversion symmetry, $Cd_3As_2$ is nominated for Dirac semimetal due to its linear band crossing at the Fermi level, and $WTe_2$ is nominated for resonant compensated semimetal due to its perfect electron-hole symmetry.[6–9] Here we introduce LaSb, a simple rock-salt structure material without broken inversion symmetry, without perfect linear band crossing, and without perfect electron-hole symmetry. Yet LaSb portrays all the exotic field induced behaviors of the aforementioned semimetals in an archetypal fashion. It shows (a) the universal TI resistivity with a plateau at $T \approx 15$ K, revealed by a magnetic field, (b) ultrahigh mobility $\mu \approx 10^5$ $cm^2V^{-1}s^{-1}$ of carriers in the plateau region, (c) quantum oscillations with a non-trivial Berry phase, and (d) XMR of about one million percent at $H = 9$ T rivaled only by $WTe_2$ and NbP. Due to its dramatic simplicity, LaSb is the ideal model **system to formulate a theoretical understanding of the exotic consequences of breaking TRS in TSMs.**

**Introduction:**

Recent advances in the band theory of solids have resulted in a novel topological classification of insulators. The theory relies on the conservation of time reversal symmetry (TRS) that leads to a robust two-fold degeneracy of bands at certain high symmetry points in the Brillouin zone known as Kramer's degeneracy. In the presence of strong spin-orbit coupling, the Kramer's pairs can interchange partners as they cross the Brillouin zone from one high symmetry point to another, and therein, the TI is born.[10–13] At the boundary between insulators of different topological classes, the band gap vanishes inevitably to allow for a change of band structure topology.[1] Therefore, the surface of a TI that is in contact with air (a trivial insulator) is metallic while the bulk is insulating. The transport signature of such combination is an initial increase of the resistivity with decreasing temperature, followed by a resistivity plateau where the metallic surface conductance saturates the insulating bulk resistance. Such *universal TI resistivity* is observed in diverse TI candidates from BTS to $SmB_6$.[2–5,14] All the exciting applications proposed for TIs rely on the robustness of their conducting surface modes protected by TRS. An intriguing question remains as to whether robust surface conduction can be induced but by breaking TRS, a regime where TIs cannot survive. In this work, we show that the answer is positive.

We present the rare-earth antimonide, LaSb, which like $SmB_6$ has a cubic structure; $SmB_6$ is simple cubic (space group P m-3m) while LaSb is face centered cubic (F m-3m). Through a series of transport experiments in a magnetic field, we arrived at two important observations: First, a field-induced resistivity activation that seems to diverge initially, but below ~ 15 K, the insulating-like divergence is arrested by a plateau (Fig. 1a, red curve). Comparing the red curves in Fig. 1a and 1b shows analogous behavior in the resistivity of TI candidates such as $SmB_6$. Our second observation is intimately connected to the first one. The resistivity of the plateau in LaSb dramatically enhances in response to a magnetic field, giving rise to an extremely large magnetoresistance (MR) of about one million percent at a moderate magnetic field $H = 9$ T (Fig. 3a, b). Finding materials with large MR has immediate applications in magnetic sensors, switches, and memories.[15–17] Traditionally, giant and colossal magnetoresistance (GMR ~ $10^2$ % and CMR ~ $10^4$ %) have been found in magnetic oxide materials.[17,18] While most non-magnetic metals have only a few percent MR, several recent studies show extreme magnetoresistance (XMR) in TaAs, NbAs, $NbSb_2$, NbP, $Cd_3As_2$, $PtSn_4$, and $WTe_2$.[6–9,19–22] LaSb has several advantages over the aforementioned semimetals: (a) It shares the record for the highest MR with NbP and $WTe_2$, (b) It does not contain Arsenic, (c) It is fairly stable to air or moisture, and (d) it has a simple rock-salt structure, ideal for theoretical calculations.

A recent theoretical work by Zeng *et al.* has predicted that LaSb is a TI with a 10 meV gap that opens near the X point of the Brillouin zone (SI-E).[23] Motivated by this work, we grew LaSb single crystals using tin flux (SI-A, Fig. 1a, *inset*). Measurements were performed using a commercial Quantum Design PPMS on two samples labeled Sample 1 (higher quality) and Sample 2. The field direction is always parallel to the [001] direction of the crystal.

**Results:**

Fig. 1a and 1b compare the normalized resistance of LaSb with the TI candidate $SmB_6$ to illustrate that the field tuning of resistivity in LaSb is analogous the pressure tuning of resistivity in $SmB_6$. Each $R(T)$ curve in Fig. 1 is normalized by its individual value at $T = 80$ K. Contrary to the theoretical prediction by Zeng *et al.*,[23] LaSb is not an insulator at zero field. Instead, it is a metal with a residual resistivity as low as 80 n$\Omega$ cm, comparable to the highest quality intermetallic compounds.[24] Fig. 1a shows that a magnetic field of $H = 9$ T activates the $R(T)$ curve into an insulator-like increase with decreasing temperature. However, this initial increase plateaus at $T \approx$ 15 K, making the $R(T)$ curve in LaSb at $H = 9$ T comparable to the $R(T)$ curve in SmB$_6$ at $H = 0$ (red curves in Fig. 1a and 1b). The plateau of LaSb onsets at $T \approx 15$ K, three times higher than $T \approx 5$ K in SmB$_6$. Fig. 1b shows that 7 GPa of pressure completely suppresses the insulating resistance of SmB$_6$ and turns it into a metal whose $R(T)$ curve is similar to LaSb at $H = 0$ (green curves in Fig. 1b and 1a). To summarize, increasing the magnetic field in LaSb and decreasing the pressure in SmB$_6$ generate the same phenomenological changes in the normalized $R(T)$. The insulator-like resistivity followed by a plateau in SmB$_6$ has been taken as a transport signature of conducting surface states of a TI protected by TRS.[4,5,25] Revealing the same plateau by *breaking* TRS in LaSb opens a new approach to engineer robust metallic surface modes.

Figs. 2a and 2b show the temperature dependence of resistivity in Sample 1 and Sample 2 at different fields. The blue curves at $H = 0$ are metallic from $T = 300$ K to 2 K. Magnetic field induces an insulator-like activation in $\rho(T)$ that grows larger as the field is increased from $H = 0$ to 9 T. The inset of Fig. 2b shows that the resistivity activation onsets at $H_{onset} = 0.4 \pm 0.1$ T, above which, $\rho(T)$ acquires a minimum at $T_m$. Fig. 2c shows the derivative $\partial\rho/\partial T$ curves for sample 1 at different fields. $T_m$ appears as the point of sign change in $\partial\rho/\partial T$ (Fig. 2c, *inset*). Figs. 2a and 2b show that the increasing resistivity of LaSb goes through an inflection and then plateaus. The temperature where $\rho(T)$ plateaus is ill-defined, instead, the inflection point which appears as a peak in $\partial\rho/\partial T$, is well-defined (Fig. 2c). We call the temperature of the inflection $T_i$ and take it as a marker of the short-circuiting induced by the surface conduction. Fig. 2d shows the evolution of $T_m$ and $T_i$ as a function of magnetic field. While $T_m$ clearly increases with increasing field, $T_i$ remains almost unchanged, showing that the extension of the insulator-like regime in temperature grows with increasing field while the extension of the plateau remains unchanged. Fig. 2d and the inset of Fig. 2b show that $T_m$ and $T_i$ merge at $H_{onset}$ implying that the resistivity plateau exists even at zero field. A similar behavior is seen in $SmB_6$ whose metallic resistivity at $P = 7$ GPa seems to contain the same plateau that exists in the insulating curve at zero pressure (Fig. 1b). Fig. 2e is a plot of $Log(\rho)$ as a function of $T^{-1}$ that allows us to estimate an insulating gap for $T < T_m$ using $\rho(T) \propto \exp(\Delta/k_B T)$. The resulting gap values are plotted as a function of field in Fig. 2f. The higher quality of Sample 1 is reflected in its higher gap magnitude at $H = 9$ T (Fig. 2f) and lower residual resistivity at $H = 0$ (Fig. 4a). A live comparison between samples of different quality is especially helpful in understanding the effect of field on LaSb.

Fig. 3 analyzes the field dependence of resistivity, showing the second important consequence of breaking TRS in LaSb: an extremely large MR $\equiv \frac{R(H)-R(0)}{R(0)}$ of order $10^6$ % in Sample 1 and $10^5$ % in sample 2 at $H = 9$ T (Fig. 3a and 3b). Notice that MR($H$) displays a non-saturating $H^2$ dependence in the higher quality Sample 1 while displaying a saturating behavior in Sample 2, reflecting a larger mean free path in Sample 1. Given comparable residual resistivity ratios (RRR), LaSb, NbP, and $WTe_2$ present the largest MR across all non-magnetic semimetals (SI-B). The ripples on the MR data at $T = 2$ K and 5 K are quantum oscillations (discussed later). Fig. 3c shows that the higher MR in Sample 1, compared to Sample 2, originates from both a higher $\rho(H = 9$ T$)$ and a lower $\rho(H = 0)$ (Fig. 3c, *inset*). Fig. 3d shows the resistance of Sample 2 (left y-axis) as a function of temperature (lower x-axis) as well as its MR (right y-axis) as a function of magnetic field (upper x-axis), before and after roughening its surface using an alumina abrasive plate. Both $R(T)$ and MR($H$) remain almost unchanged after surface abrasion. Similar experiments have been performed on $SmB_6$ by Kim *et al.* to prove the robustness of surface states of the TI.[4] We use resistance and not resistivity in Fig. 3d to eliminate geometric errors due to surface abrasion. MR, by definition, is independent of geometric factors.

To determine the sign and the mobility of the carriers in LaSb, we use a combination of resistivity and Hall effect. Fig. 4a shows a power law fit of the form $\rho = \rho_0 + A\,T^4$, below $T = 20$ K, to extract $\rho_0 = 80$ nΩ cm in Sample 1 and $\rho_0 = 330$ nΩ cm in Sample 2. Fig. 4b shows the temperature dependence of the Hall coefficient $R_H(T)$ at $H = 9$ T in both samples. $R_H(T)$ is positive and small for 40 K $< T <$ 200 K, but negative and large for $T <$ 40 K, showing partial compensation between p-type and n-type carriers with the latter being dominant at lower temperatures. Using the relation $n = \frac{1}{eR_H(0)}$ where $R_H(0)$ is the zero temperature limit of $R_H(T)$ (Fig. 4b, *inset*), we estimate the n-type carrier concentration $n = 1.1 \times 10^{20}$ cm$^{-3}$ in Sample 1 and $n = 1.7 \times 10^{20}$ cm$^{-3}$ in Sample 2. Using the relation $\mu_H = \frac{R_H(0)}{\rho_0}$, we estimate the Hall mobility $\mu_H = 4.4 \times 10^5$ cm$^2$V$^{-1}$s$^{-1}$ in Sample 1 and $\mu_H = 1.7 \times 10^5$ cm$^2$V$^{-1}$s$^{-1}$ in Sample 2. The high mobility of the n-type carriers results in the Shubnikov-de Haas (SdH) oscillations shown in Fig. 4c at $T = 2$ K and 5 K in the range 5 T $< H <$ 9 T. After subtracting a smooth background (solid black line in Fig. 4c), the oscillations are plotted as a function of $H^{-1}$ for $\frac{1}{9\,T} < \frac{1}{H} < \frac{1}{5\,T}$ in Fig. 4d. Fourier transforms of the data in Fig. 4d are shown in Fig. 4e, revealing a small frequency $F_\alpha = 212 \pm 5$ T with its harmonic $F_{2\alpha} \approx 414$ T, and a second frequency at $F_\beta = 433 \pm 5$ T. SdH oscillations confirm a metallic ground state in the plateau region. They can also be used to detect a Berry phase which is present when the Bloch states of a metal wind around contours that contain band crossings such as the Dirac points in Graphene or topological insulators.[26–29] The Onsager relation in two dimensions $\frac{\hbar}{e}\frac{A_n}{H} = 2\pi(n - \beta)$ relates the Berry phase $2\pi\beta$ to the frequency of oscillations $F = \frac{\phi_0}{2\pi^2}A_n$ where $A_n$ is the extremal orbit area of the $n^{\text{th}}$ Landau level (LL) and $\phi_0$ is the quantum of flux.[26] Fig. 4f is a plot of the LL index $n$ as a function of $H^{-1}$ where $H$ is the field at which resistivity is a maximum for integer $n$ and minimum for half-integer $n$ (SI-C and D). Our data are limited to a maximum field of $H = 9$ T, therefore we could count only down to $n = 25$. Linear fits in Fig. 4d intersect the y-axis at $n = 0.78 \pm 0.05$ indicating a non-zero Berry phase in LaSb. The slope of these linear fits must equal the frequency of oscillations according to Onsager relation. We obtain $F_{min} = 218$ T and $F_{max} = 216$ T from the lines of minima and maxima (Fig 4f), in agreement with $F_\alpha = 212$ T (Fig. 4e). The possibility of SdH oscillations originating from tin inclusions is excluded for two reasons: (a) lack of tin superconducting transition at 3.7 K (Fig. 4a, SI-A) (b) lack of the known oscillation frequencies of tin.[30]

**Discussion**

By comparing the effect of field on LaSb and the effect of pressure on $SmB_6$ (Fig. 1) we formulate a direct link between topological insulators in the limit of preserved TRS and topological semimetals (TSM) in the limit of broken TRS. The only known field induced topological metallic state is the Weyl semimetal (WSM) with Fermi arcs on its surface that could result in quantum oscillations.[31–33] This is consistent with SdH oscillations and the finite Berry phase in LaSb. The data in Fig. 2 raises an important question: "how does a WSM in the plateau region connect to the field-induced insulator-like region $T_i < T < T_m$?" Fig. 2d shows that the extension of the metallic plateau (in temperature) is unaffected by magnetic field while the extension of the insulator-like regime increases with increasing field. This suggests a scenario of two-channel conduction: (i) the plateau channel that is not sensitive to magnetic field, and (ii) the insulator-like channel that is highly sensitive to magnetic field. Further work is required to understand the link between the resistivity plateau and the field revealed insulator-like regime in LaSb and other TSMs.

The second consequence of breaking TRS in LaSb is the extreme MR (XMR) that surfs on the resistivity plateau. Recently, several TSMs have been discovered with XMR, high carrier mobility and quantum oscillations.[6–9,19–21] A range of exotic states have been proposed for these TSMs including Weyl semimetal (WSM) in NbP and TaAs due to the lack of inversion symmetry, Dirac semimetal (DSM) in $Cd_3As_2$ due to linear band crossing, and resonant compensated semimetal due to prefect electron-hole symmetry in $WTe_2$. Surprisingly, LaSb shows the same field induced properties yet its simple FCC structure does not break inversion symmetry, its band structure does not have a linear crossing (SI-E), and its electron-hole compensation is not perfect (SI-E).

The live comparison between our higher quality Sample 1 and lower quality Sample 2 makes it clear that the crystal quality, or the carrier mobility, is the key player in boosting XMR (SI-B).[22] Note that the XMR in LaSb and other TSMs is a consequence of a metal to insulator-like transition at $T_m$ induced by the magnetic field (Fig. 2a-d). The inset of Fig. 2b shows that XMR onsets at $H_{onset} = 0.4$ T suggesting potential application of LaSb as a magnetic switch and a magnetic sensor especially since there is a current strong demand for magnetic sensors in the low field low temperature regime.[34] The exotic field effects recently observed in TSMs seems to shift the focus of the topological research community from the limit of protected TRS to the limit of broken TRS. LaSb with its simple rock salt structure offers an ideal model system for detailed theoretical understanding of the field induced properties of TSMs.

**Figure Captions:**

**Figure 1 | Comparing the pressure tuned phenomenology of SmB$_6$ with the field tuned phenomenology of LaSb. a,** Normalized resistance of LaSb at $H$ = 0 (green) and $H$ = 9 T (red) plotted as a function of temperature on a log-log scale. Each $R(T)$ curve is normalized by its individual resistance at $T$ = 80 K. While the green $H$ = 0 T curve shows metallic behavior, the red $H$ = 9 T curve shows an insulator-like behavior. The increasing resistance of the red curve is arrested by a plateau at $T \approx$ 15 K marked by a black arrow. A single crystal of the cubic LaSb is shown in the inset. **b,** Normalized Resistance of SmB$_6$, adopted from Ref.[35], at $P$ = 7 GPa (green) and $P$ = 0 GPa (red) plotted as a function of temperature on a log-log scale. Each curve is normalized by its individual resistance at $T$ = 80 K. The red curve at $P$ = 0 GPa in SmB$_6$ and the red curve at $H$ = 9 T in LaSb are analogous. The plateau appears at $T \approx$ 5 K in the former and $T \approx$ 15 K in the latter. The green curve at $P$ = 7 GPa in SmB$_6$ and the green curve at $H$ = 0 T in LaSb are also analogous, both showing a metallic behavior with an arguably persistent plateau.

**Figure 2 | Temperature dependence of resistivity in LaSb. a,** Resistivity of Sample 1 plotted as a function of temperature at $H$ = 0, 1, 3, 6, and 9 T. The black arrow marks $T_m$ for the black curve at H = 9 T. **b,** Similar $\rho(T)$ curves from Sample 2 that show a lower plateau resistivity at $H$ = 9 T compared to the higher quality Sample 1. The inset shows that the activation of $\rho(T)$ onsets at $H_{onset}$ = 0.4 ± 0.1 T. **c,** $\partial\rho/\partial T$ plotted as a function of temperature for Sample 1. The peak in the $\partial\rho/\partial T$ marks the inflection point $T_i$ at each field. The sign change in the $\partial\rho/\partial T$ marks the resistivity minimum at $T_m$. The inset shows $T_m$ and $T_i$ at $H$ = 1 T. **d,** $T_m$ and $T_i$ plotted as a function of magnetic field. $T_i$ stays almost flat while $T_m$ clearly evolves with field. As the field is decreased from 9 T to zero, the two temperature scales seem to merge. **e,** $Log(\rho)$ plotted as a function of $T^{-1}$ to extract the activation gap in LaSb. The gray lines show the region of the linear fits. **f,** The resulting gap values from Fig. 2e plotted as a function of magnetic field for both samples. The higher quality Sample 1 shows higher gap values with less saturation at higher fields.

**Figure 3 | Field dependence of resistivity in LaSb. a,** Magnetoresistance $\text{MR}(\%) = 100 \times \frac{R(H)-R(0)}{R(0)}$ plotted as a function of field in Sample 1 at different temperatures. The curves at $T = 2$ K and 5 K are in the plateau region with MR of about one million percent. The wiggles in the data are quantum oscillations. MR of Sample 1 shows no sign of saturation up to $H = 9$ T. **b,** Similar plot for MR versus $H$ in the lower quality Sample 2. The magnitude of MR in Sample 2 is one order of magnitude smaller and it seems to saturate at $H = 9$ T. **c,** Resistivity in units of µΩ cm plotted as a function of magnetic field for both samples. The inset zooms into the low field region. **d,** The temperature dependence of resistance (lower x-axis, left y-axis) as well as the field dependence of magnetoresistance (upper x-axis, right y-axis) are compared before and after surface abrasion on Sample 2. No significant changes are observed.

**Figure 4 | Hall effect and quantum oscillations in LaSb. a,** Residual resistivity of Sample 1 and Sample 2 extracted from power law fits of the form $\rho = \rho_0 + A\,T^4$ to the low temperature $\rho(T)$ data. **b,** Hall effect plotted as a function of temperature in both samples showing a small positive signal at $T > 40$ K and a larger negative signal at $T < 40$ K. The inset zooms into the low temperature data. **c,** Quantum oscillations observed at $T = 2$ K and 5 K in the resistivity plateau. The data in panels c, d, e, and f are from Sample 1. A smooth background (solid black line) is subtracted from data to extract the purely oscillatory part of $\rho(H)$. **d**, The purely oscillatory Δ$\rho$ plotted as a function of inverse field. The amplitude of oscillations does not decrease monotonically implying the existence of more than one frequency. **e,** Fast Fourier Transform (FFT) of the data in panel e showing $F_\alpha = 212$ T, $F_{2\alpha} = 414$ T, and $F_\beta = 433$ T with an almost two fold amplitude reduction as the temperature is increased from $T = 2$ K to 5 K. **f,** Landau level indices plotted as a function of the field. The peaks in $\Delta\rho(H)$ correspond to integer indices (magenta circles) and the dips correspond to half-integer indices (green circles). The slope of the two linear fits are consistent with the principle frequency $F_\alpha = 212$ T.

**References:**

1. Moore, J. E. The birth of topological insulators. *Nature* **464,** 194–198 (2010).

2. Ren, Z., Taskin, A. A., Sasaki, S., Segawa, K. & Ando, Y. Large bulk resistivity and surface quantum oscillations in the topological insulator $Bi_2Te_2Se$. *Phys. Rev. B* **82,** 241306 (2010).

3. Jia, S. *et al.* Defects and high bulk resistivities in the Bi-rich tetradymite topological insulator $Bi_{2+x}Te_{2-x}Se$. *Phys. Rev. B* **86,** 165119 (2012).

4. Kim, D. J. *et al.* Surface Hall Effect and Nonlocal Transport in $SmB_6$: Evidence for Surface Conduction. *Sci. Rep.* **3,** (2013).

5. Kim, D. J., Xia, J. & Fisk, Z. Topological surface state in the Kondo insulator samarium hexaboride. *Nat. Mater.* **13,** 466–470 (2014).

6. Yang, L. *et al.* Discovery of a Weyl Semimetal in non-Centrosymmetric Compound TaAs. *ArXiv150700521 Cond-Mat* (2015).

7. Shekhar, C. *et al.* Extremely large magnetoresistance and ultrahigh mobility in the topological Weyl semimetal NbP. *Nat. Phys.* (2015).

8. Liang, T. *et al.* Ultrahigh mobility and giant magnetoresistance in the Dirac semimetal $Cd_3As_2$. *Nat. Mater.* **14,** 280–284 (2015).

9. Ali, M. N. *et al.* Large, non-saturating magnetoresistance in $WTe_2$. *Nature* **514,** 205–208 (2014).

10. Ando, Y. Topological Insulator Materials. *J. Phys. Soc. Jpn.* **82,** 102001 (2013).

11. Hasan, M. Z. & Kane, C. L. *Colloquium*: Topological insulators. *Rev. Mod. Phys.* **82,** 3045–3067 (2010).

12. Kane, C. L. & Mele, E. J. $Z_2$ Topological Order and the Quantum Spin Hall Effect. *Phys. Rev. Lett.* **95,** 146802 (2005).

13. Fu, L., Kane, C. L. & Mele, E. J. Topological Insulators in Three Dimensions. *Phys. Rev. Lett.* **98,** 106803 (2007).

14. Dzero, M., Sun, K., Galitski, V. & Coleman, P. Topological Kondo Insulators. *Phys. Rev. Lett.* **104,** 106408 (2010).

15. Daughton, J. M. GMR applications. *J. Magn. Magn. Mater.* **192,** 334–342 (1999).

16. Lenz, J. E. A review of magnetic sensors. *Proc. IEEE* **78,** 973–989 (1990).

17. Moritomo, Y., Asamitsu, A., Kuwahara, H. & Tokura, Y. Giant magnetoresistance of manganese oxides with a layered perovskite structure. *Nature* **380,** 141–144 (1996).

18. Ramirez, A. P., Cava, R. J. & Krajewski, J. Colossal magnetoresistance in Cr-based chalcogenide spinels. *Nature* **386,** 156–159 (1997).

19. Ghimire, N. J. *et al.* Magnetotransport of single crystalline NbAs. *J. Phys. Condens. Matter* **27,** 152201 (2015).

20. Wang, K., Graf, D., Li, L., Wang, L. & Petrovic, C. Anisotropic giant magnetoresistance in $NbSb_2$. *Sci. Rep.* **4,** (2014).

21. Mun, E. *et al.* Magnetic field effects on transport properties of $PtSn_4$. *Phys. Rev. B* **85,** 035135 (2012).

22. Ali, M. N. *et al.* Correlation of crystal quality and extreme magnetoresistance of $WTe_2$. *EPL Europhys. Lett.* **110,** 67002 (2015).

23. Zeng, M. *et al.* Topological semimetals and topological insulators in rare earth monopnictides. *ArXiv150403492 Cond-Mat* (2015).

24. Saxena, S. S. *et al.* Superconductivity on the border of itinerant-electron ferromagnetism in UGe$_2$. *Nature* **406,** 587–592 (2000).

25. Wolgast, S. *et al.* Low-temperature surface conduction in the Kondo insulator SmB$_6$. *Phys. Rev. B* **88,** 180405 (2013).

26. Mikitik, G. P. & Sharlai, Y. V. Manifestation of Berry's Phase in Metal Physics. *Phys. Rev. Lett.* **82,** 2147–2150 (1999).

27. Wright, A. R. & McKenzie, R. H. Quantum oscillations and Berry's phase in topological insulator surface states with broken particle-hole symmetry. *Phys. Rev. B* **87,** 085411 (2013).

28. Novoselov, K. S. *et al.* Two-dimensional gas of massless Dirac fermions in graphene. *Nature* **438,** 197–200 (2005).

29. Zhang, Y., Tan, Y.-W., Stormer, H. L. & Kim, P. Experimental observation of the quantum Hall effect and Berry's phase in graphene. *Nature* **438,** 201–204 (2005).

30. Deacon, J. M. & Mackinnon, L. Ultrasonic quantum oscillations in white tin. *J. Phys. F Met. Phys.* **3,** 2082 (1973).

31. Witczak-Krempa, W., Chen, G., Kim, Y. B. & Balents, L. Correlated Quantum Phenomena in the Strong Spin-Orbit Regime. *Annu. Rev. Condens. Matter Phys.* **5,** 57–82 (2014).

32. Tafti, F. F., Ishikawa, J. J., McCollam, A., Nakatsuji, S. & Julian, S. R. Pressure-tuned insulator to metal transition in Eu$_2$Ir$_2$O$_7$. *Phys. Rev. B* **85,** 205104 (2012).

33. Potter, A. C., Kimchi, I. & Vishwanath, A. Quantum oscillations from surface Fermi arcs in Weyl and Dirac semimetals. *Nat. Commun.* **5,** (2014).

34. Jankowski, J., El-Ahmar, S. & Oszwaldowski, M. Hall sensors for extreme temperatures. *Sensors* **11,** 876–885 (2011).

35. Gabáni, S. *et al.* Pressure-induced Fermi-liquid behavior in the Kondo insulator SmB$_6$ Possible transition through a quantum critical point. *Phys. Rev. B* **67,** 172406 (2003).

36. Zhu, Z. *et al.* Quantum Oscillations, Thermoelectric Coefficients, and the Fermi Surface of Semimetallic WTe$_2$. *Phys. Rev. Lett.* **114,** 176601 (2015).

**Acknowledgements:**

This research was supported by the Gordon and Betty Moore Foundation under the EPiQS program, grant GBMF 4412. S.K. is supported by the ARO MURI on topological insulators, grant W911NF-12-1-0461.

**Figure 1**

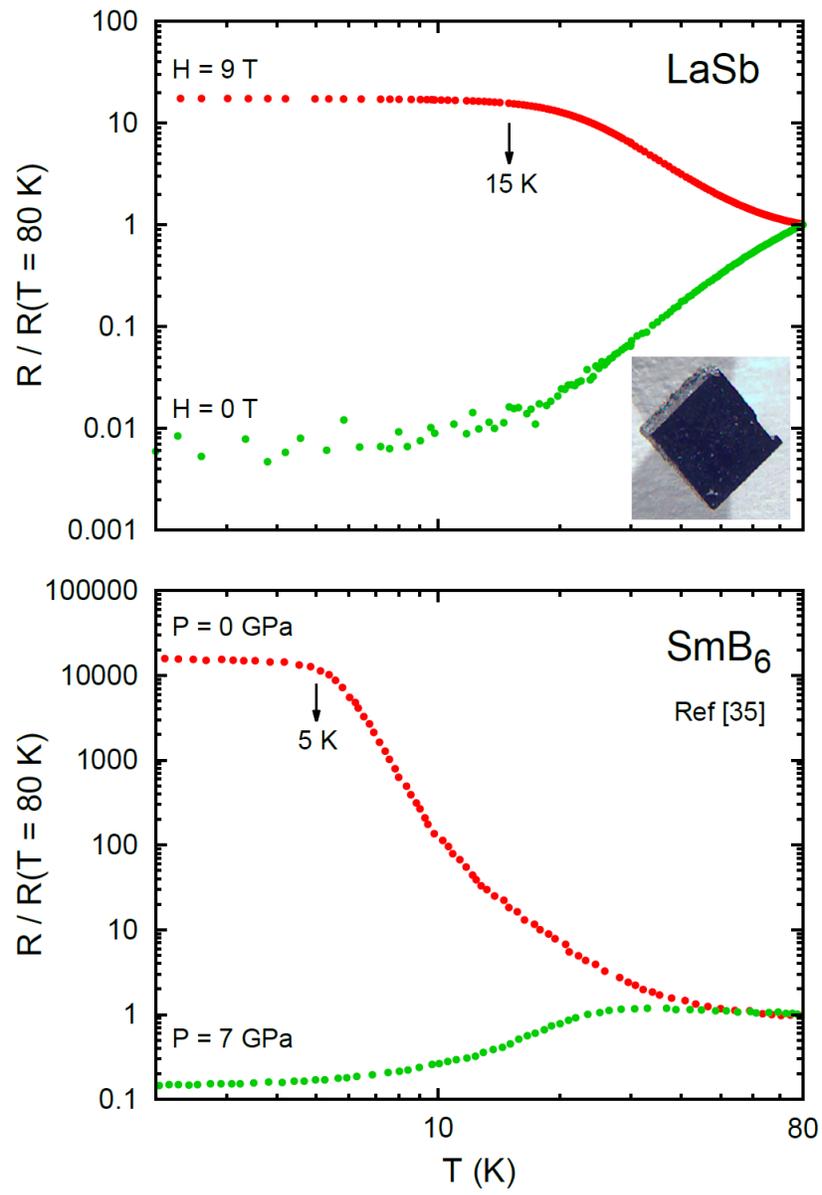

**Figure 2**

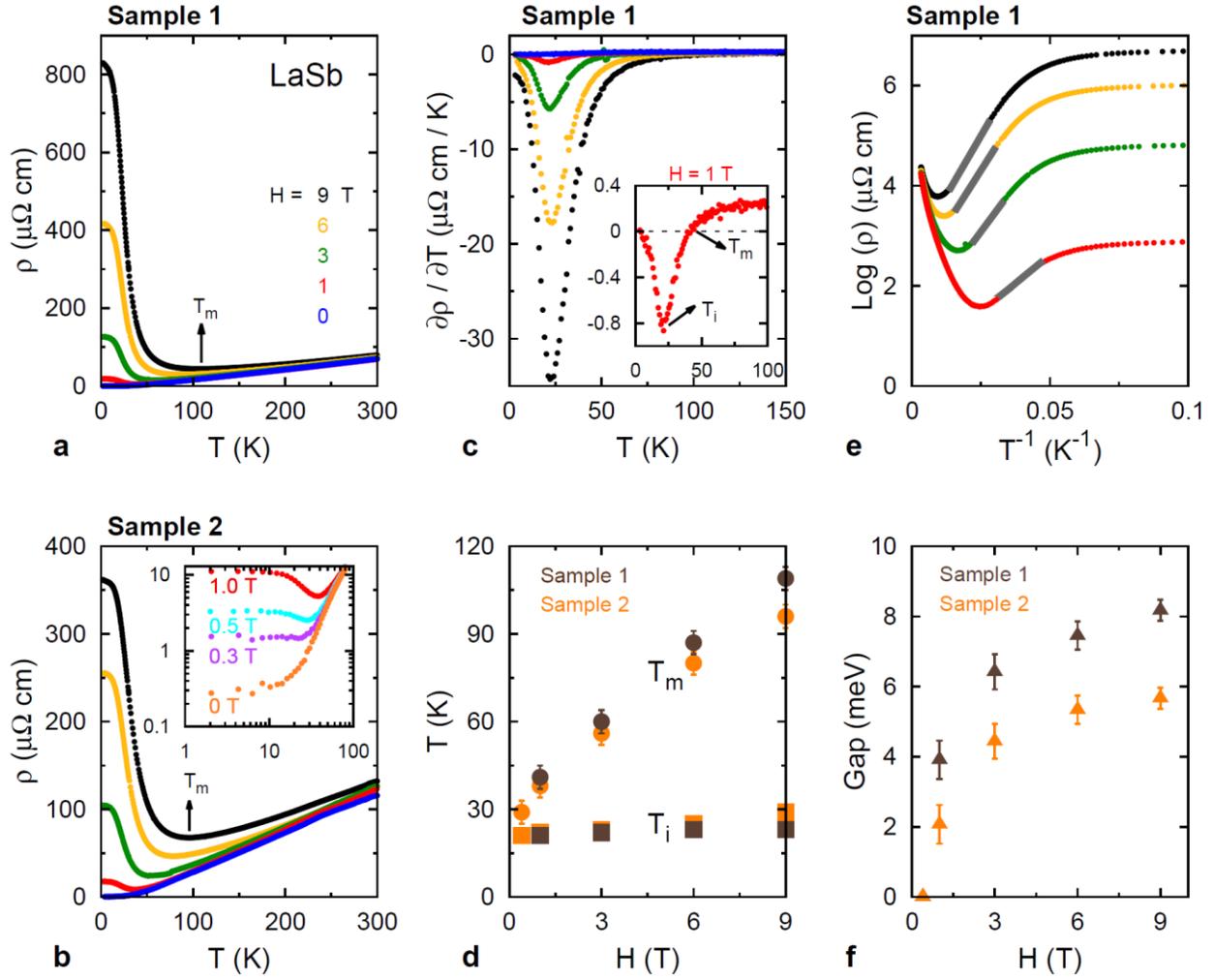

**Figure 3**

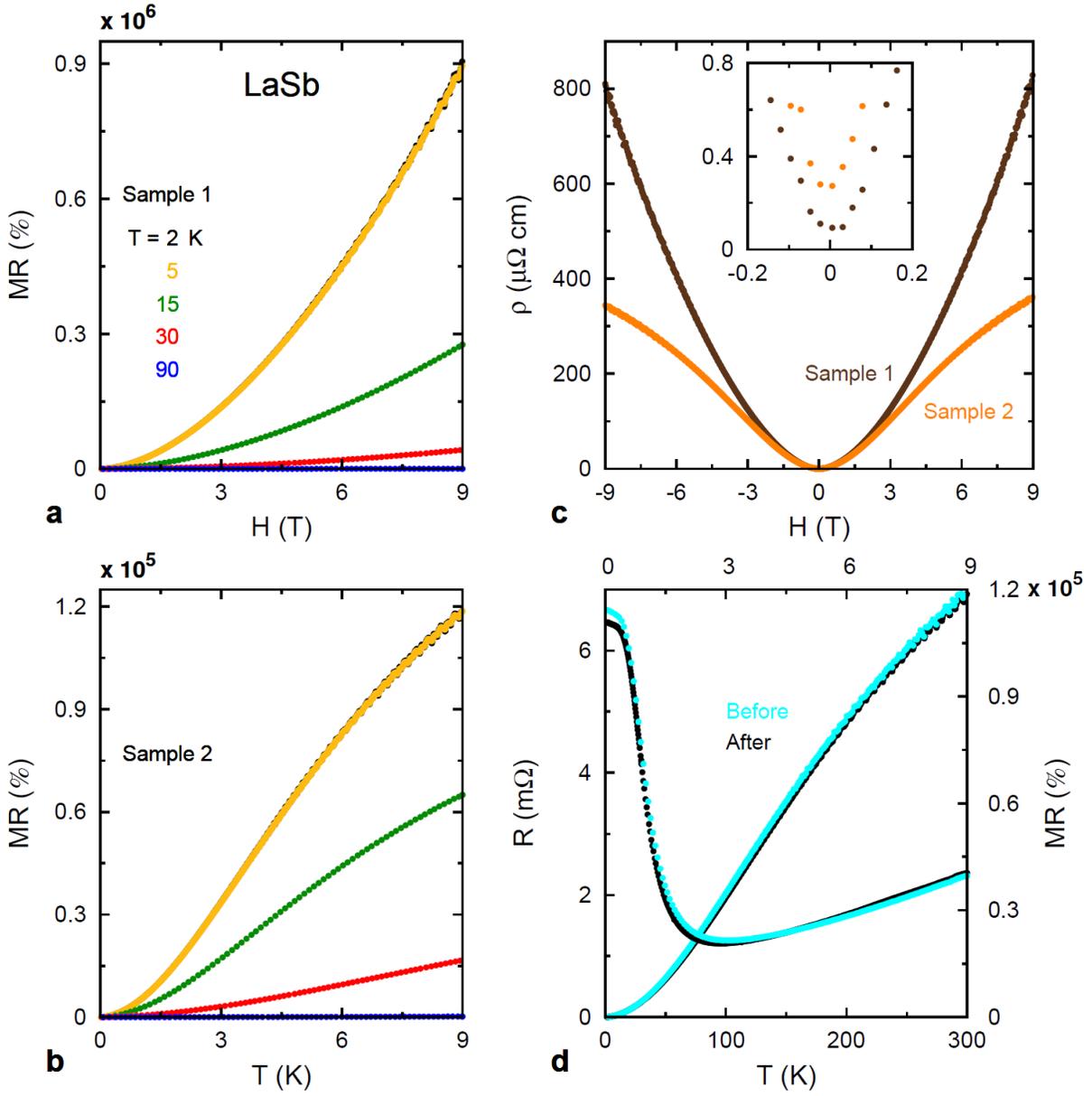

**Figure 4**

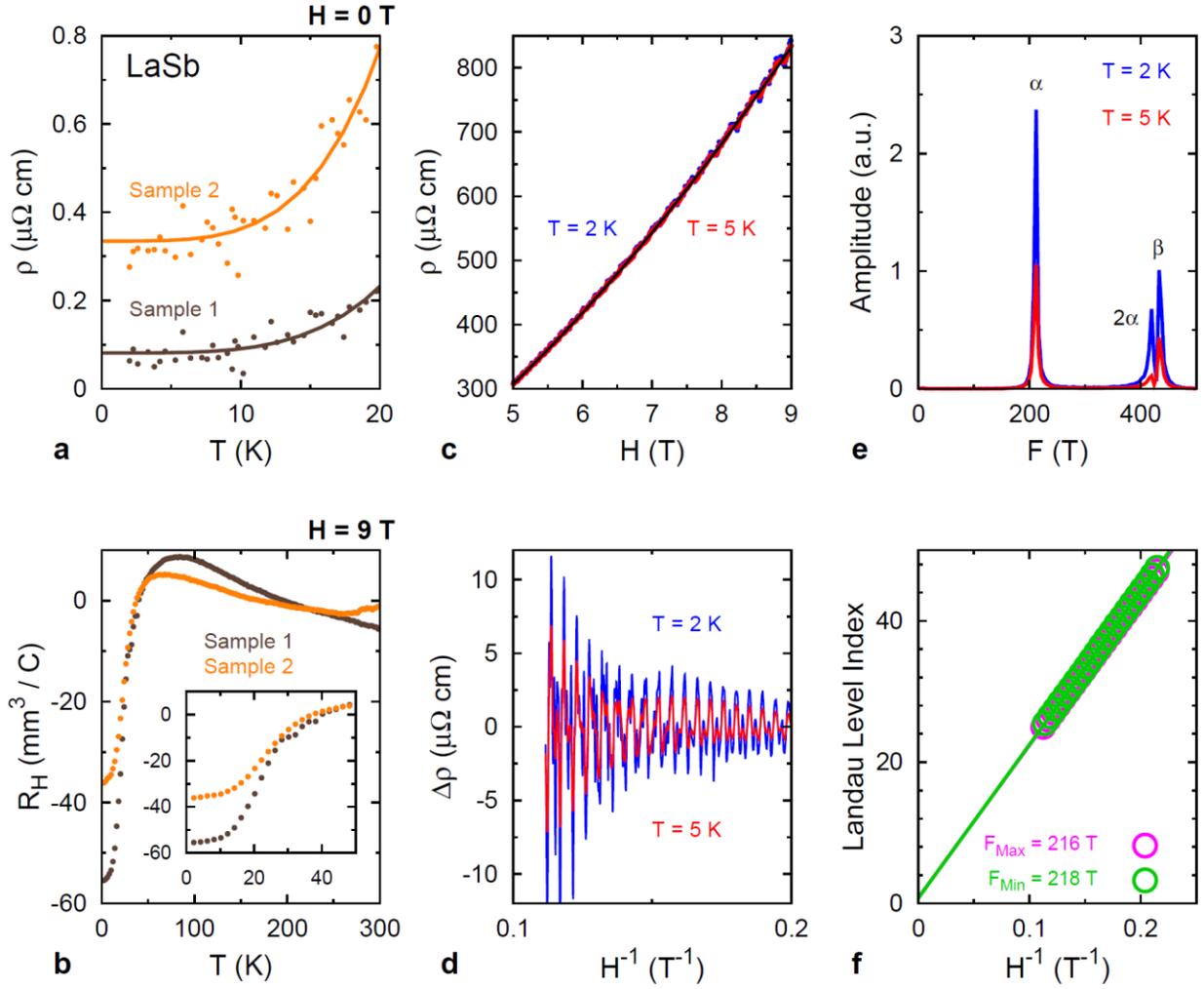

# Supplementary Information

**Section A: Sample preparations**

Single crystals of LaSb were grown out of tin flux by placing La:Sb:Sn = 1:1:20 in an alumina crucible sealed in an evacuated quartz tube, heated to 1000 C, slowly cooled to 700 C, and finally canted with a centrifuge. Sample 1 with a higher quality was grown by choosing a different ratio of La:Sb:Sn = 2:1:40. The higher quality of Sample 1 is reflected in its lower residual resistivity at zero field (Fig. 4a), higher MR at $H = 9T$ (Fig. 3c), higher gap (Fig. 2f), and higher RRR (Fig. S1). Our live comparison between samples of different quality shows how increasing sample quality drastically changes the magnitude of magnetoresistance. Fig. S1 shows resistivity data at zero field from $T = 300$ K to 2 K with the RRR values extracted from a power law fit. The lack of a superconducting transition at 3.5 K excludes tin inclusion in our samples.

**Section B: XMR in various semimetals**

Recent efforts in growing crystals of compensated semimetals have given rise to a number of intermetallic compounds with extremely large magnetoresistance (XMR). Table S1 compares RRR, carrier mobility, and MR among these compounds. LaSb, NbP, and WTe$_2$ by far have the largest XMR, given comparable crystal quality.

**Table S1** | Residual resistivity ratio (RRR), magnetoresistance (MR) at T = 2 K and H = 9 T, and mobility of carriers (μ) is compared between different semimetals that show XMR.

| Material | RRR | MR (%) | μ (cm$^2$V$^{-1}$s$^{-1}$) | Reference |
|---|---|---|---|---|
| LaSb | 346 | $1.2 \times 10^5$ | $1.7 \times 10^5$ | This work |
|  | 875 | $0.9 \times 10^6$ | $4.4 \times 10^6$ | This work |
| WTe$_2$ | 370 | $1.7 \times 10^5$ | $1.0 \times 10^5$ | Ali et al.[9] |
|  | 1256 | $2.5 \times 10^5$ | ? | Zhu et al.[36] |
| NbAs | 72 | $2.3 \times 10^5$ | $3.5 \times 10^5$ | Ghimire et al.[19] |
| TaAs | 15 | $8.0 \times 10^4$ | $1.8 \times 10^5$ | Yang et al.[6] |
| NbP | 150 | $8.5 \times 10^5$ | $5.0 \times 10^6$ | Shekhar et al.[7] |
| Cd$_3$As$_2$ | 781 | $5.8 \times 10^4$ | $3.0 \times 10^6$ | Liang et al.[8] |
|  | 4100 | $1.3 \times 10^5$ | $8.7 \times 10^6$ | Liang et al.[8] |

**Section C: Comparing the transverse and the longitudinal resistivity in LaSb**

The conductivity tensor is defined as: $\sigma_{ij} = \frac{1}{\rho_{xx}\rho_{yy}-\rho_{xy}\rho_{yx}}\begin{pmatrix} \rho_{yy} & -\rho_{xy} \\ -\rho_{yx} & \rho_{xx} \end{pmatrix}$. As a result of the cubic symmetry in LaSb $\rho_{xx} = \rho_{yy}$ and $\rho_{yx} = \rho_{xy}$. Using $\sigma = 1/\rho_{xx}$ is justified only when the ratio $\rho_{xy}/\rho_{xx}$ is small. Fig. S2 compares the magnitude of the longitudinal ($\rho_{xx}$) versus the transverse ($\rho_{xy}$) resistivity in Sample 1 and 2. Due to the small $\rho_{xy}/\rho_{xx}$ ratio, we can safely use the maxima and the minima of quantum oscillations in the $\rho_{xx}$ channel for our analysis of the Berry phase in Fig. 4 and for the extraction of the semiconducting gap in Fig. 2.

**Section D: Indexing the Landau levels from quantum oscillations in resistivity**

We used Shubnikov-de Haas (SdH) oscillations to detect a Berry phase in LaSb. Every time the Fermi level falls between two Landau levels, corresponding to an integer number of filled Landau levels, conductivity shows a minimum and therefore resistivity shows a maximum. As explained in SI. C and Fig. S2, $\rho_{xx} = 1/\sigma$ for LaSb due the small $\rho_{xy}/\rho_{xx}$ ratio. Fig. S3 shows how we index the landau levels from the resistivity data. Fig. 4f in the main text, plots these indices as a function of the corresponding $1/H$ values.

**Section E: DFT calculation on LaSb**

Fig. S4 shows the results of our density functional theory (DFT) computation using WIEN2K code. In agreement with Zeng *et al.*, we observe two hole-like Antimony bands at Γ, and one electron-like band at X with a small gap near the X point where La and Sb states invert (the latter plotted as fat bands). A linear dispersion exists near the avoided crossing at X. The size of the electron and the hole pockets are not identical, hence, LaSb is not a resonant compensated semimetal, different from WTe$_2$. Note that using an MBJ functional gives rise to a different result in LaSb where the gap is larger, the dispersion is more quadratic, and there is no band inversion. It is important to notice that band structure calculations are sensitive to the choice of the functional.

**Figure S1**

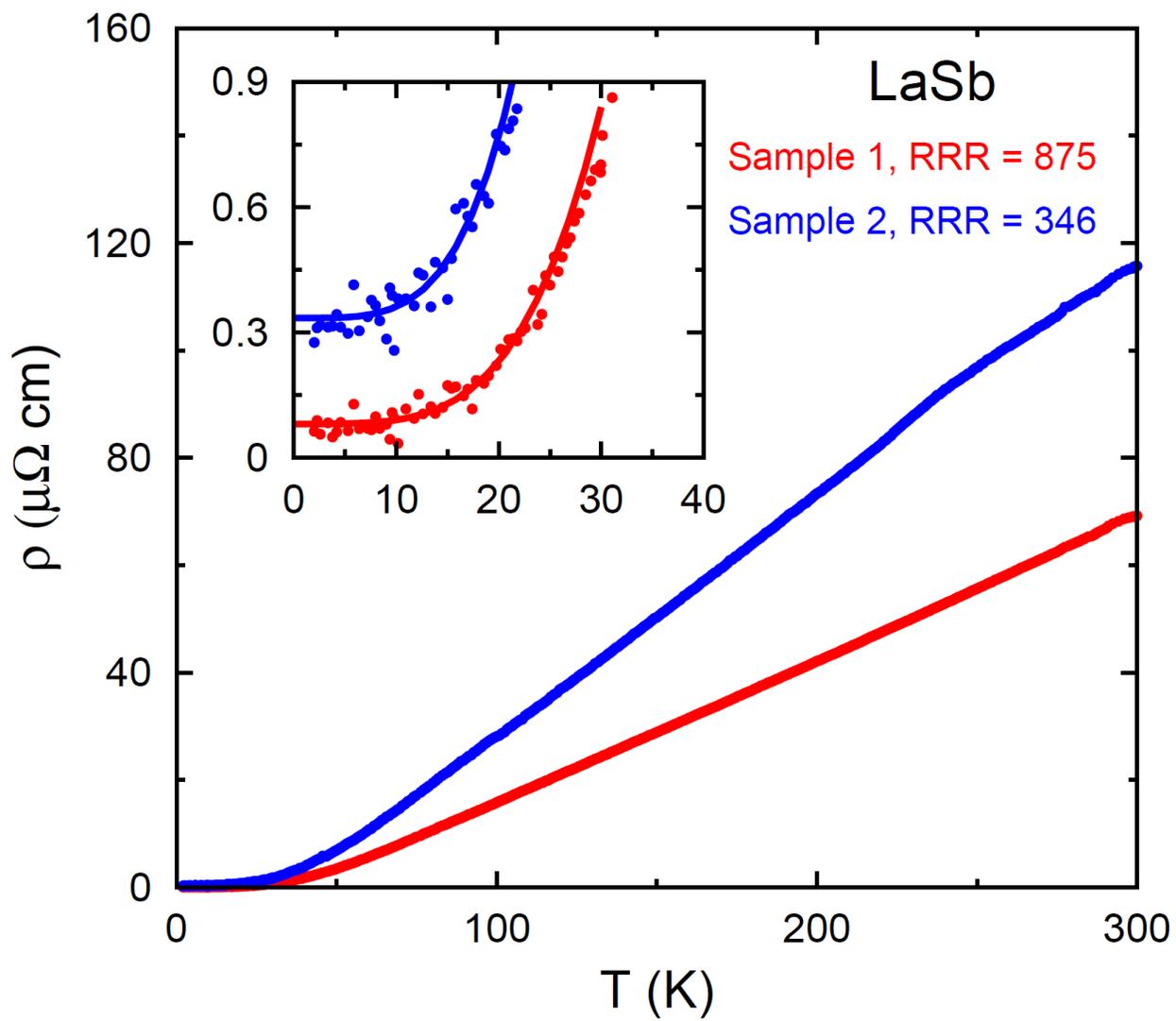

**Figure S2**

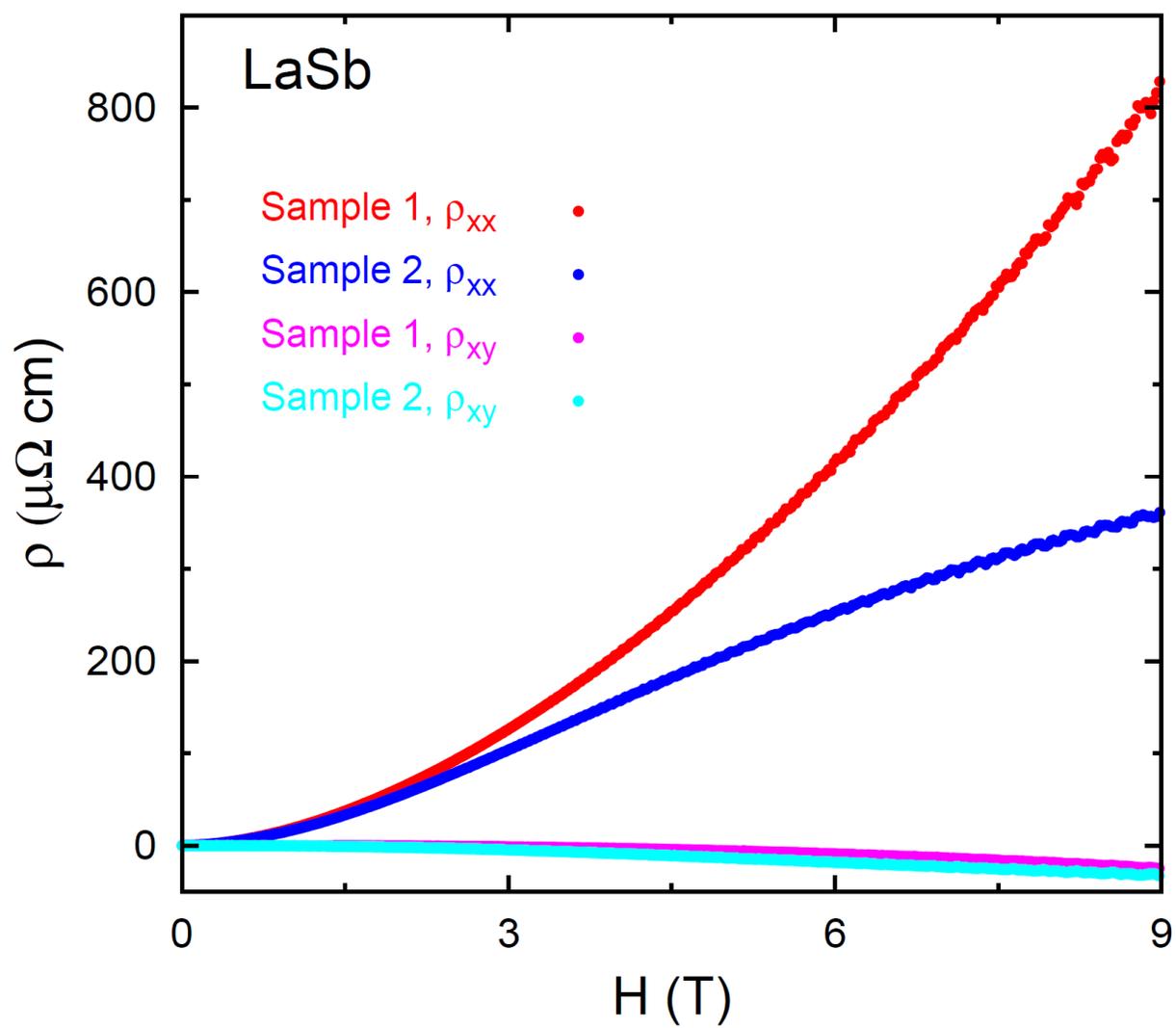

**Figure S3**

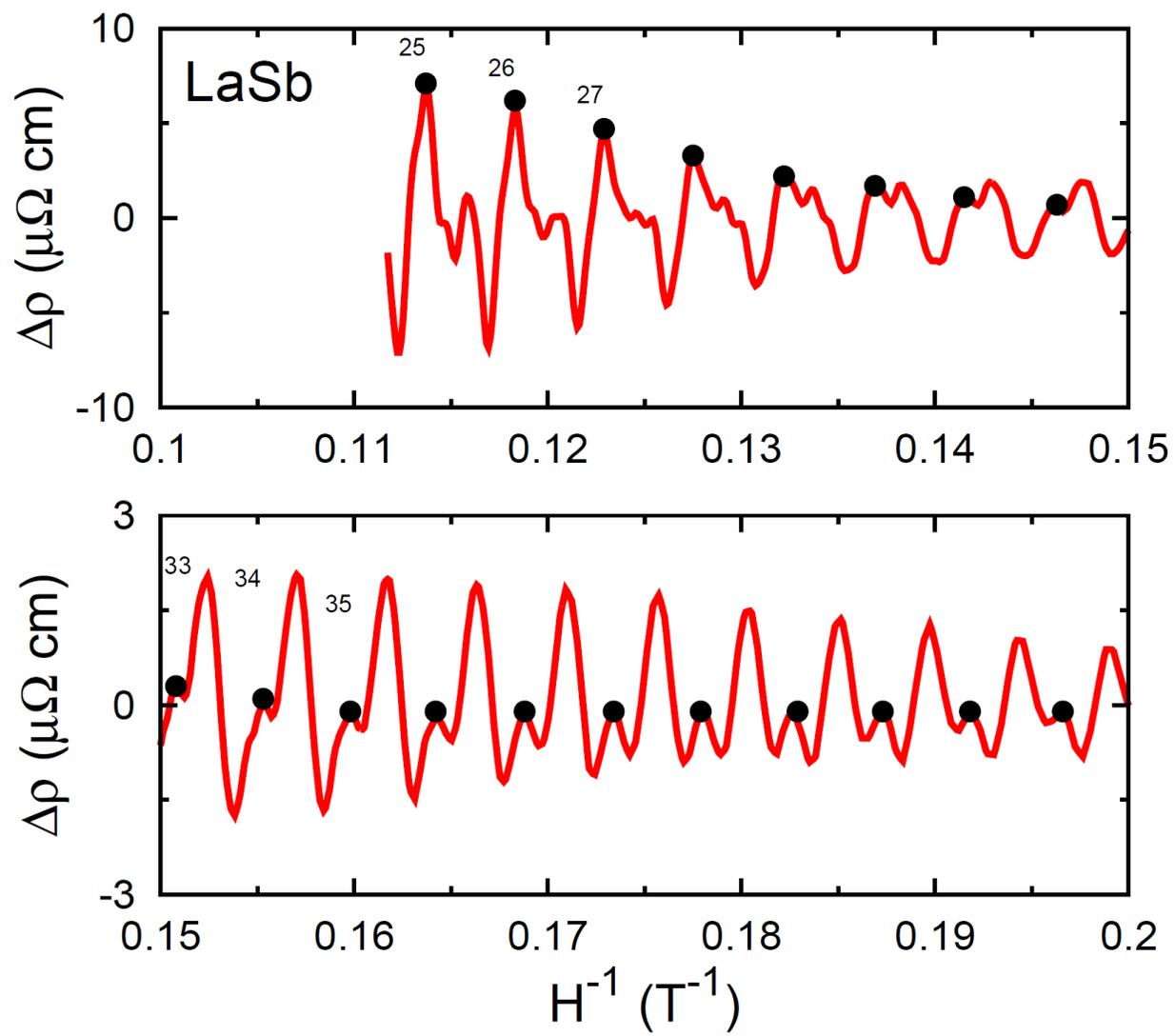

**Figure S4**

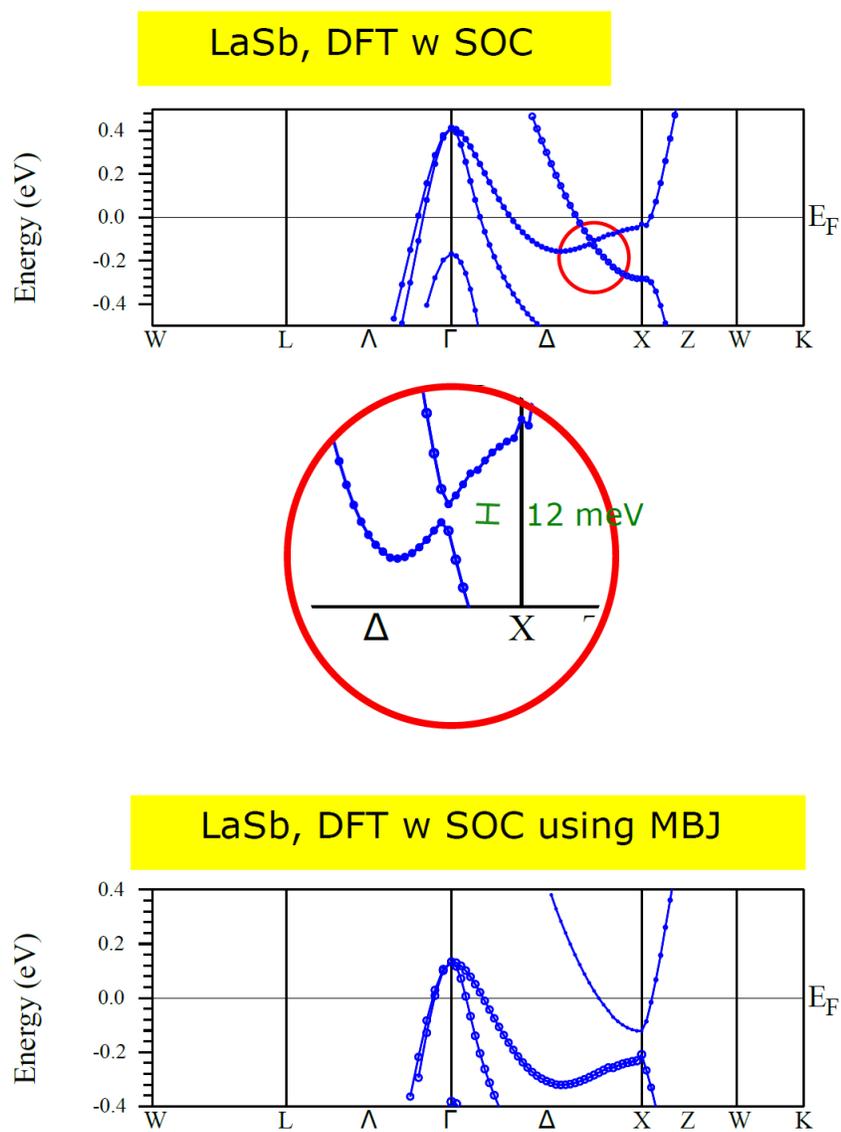